\documentclass[aps,prc,showpacs,twocolumn,superscriptaddress]{revtex4}

\usepackage{graphics}
\usepackage{latexsym}

\begin{document}


\title{Non-statistical fluctuations for deep inelastic processes\\ 
 in 
$^{27}$Al + $^{27}$Al
collision}


\author{I. Berceanu}
\affiliation{Institute of Physics and Nuclear Engineering, P.O.Box MG-6,
76900 Bucharest}
\author{M. Duma}
\affiliation{Institute of Physics and Nuclear Engineering, P.O.Box MG-6,
76900 Bucharest}
\author{D. Mois\u a}
\affiliation{Institute of Physics and Nuclear Engineering, P.O.Box MG-6,
76900 Bucharest}
\author{M. Petrovici}
\affiliation{Institute of Physics and Nuclear Engineering, P.O.Box MG-6,
76900 Bucharest}
\author{A. Pop}
\affiliation{Institute of Physics and Nuclear Engineering, P.O.Box MG-6,
76900 Bucharest}
\author{V. Simion}
\affiliation{Institute of Physics and Nuclear Engineering, P.O.Box MG-6,
76900 Bucharest}
\author{A. Del Zoppo}
\affiliation{Istituto Nazionale di Fisica Nucleare, Laboratorio Nazionale 
del Sud, v. S. Sofia 44, I-95100, Catania, Italy}
\author{G. d'Erasmo}
\affiliation{Istituto Nazionale di Fisica Nucleare, Sezione di Bari, v. Amendola 173, 70125 Bari, Italy}
\affiliation{Dipartimento di Fisica, Universita di Bari, Italy}
\author{G. Imm\' e}
\affiliation{Istituto Nazionale di Fisica Nucleare, Laboratorio Nazionale 
del Sud, v. S. Sofia 44, I-95100, Catania, Italy}
\affiliation{Dipartimento  di Fisica, Universita di Catania, I-95129 Catania, 
Italy}
\author{G. Lanzan\` o}
\affiliation{Istituto Nazionale di Fisica Nucleare, Laboratorio Nazionale 
del Sud, v. S. Sofia 44, I-95100, Catania, Italy}
\affiliation{Dipartimento  di Fisica, Universita di Catania, I-95129 Catania, 
Italy}
\author{A. Pagano}
\affiliation{Istituto Nazionale di Fisica Nucleare, Laboratorio Nazionale 
del Sud, v. S. Sofia 44, I-95100, Catania, Italy}
\affiliation{Dipartimento  di Fisica, Universita di Catania, I-95129 Catania, 
Italy}
\author{A. Pantaleo}
\affiliation{Istituto Nazionale di Fisica Nucleare, Sezione di Bari, v. Amendola 173, 70125 Bari, Italy}
\author{G. Raciti}
\affiliation{Istituto Nazionale di Fisica Nucleare, Laboratorio Nazionale 
del Sud, v. S. Sofia 44, I-95100, Catania, Italy}
\affiliation{Istituto Nazionale di Fisica Nucleare, Sezione di Catania, 
 Catania, Italy}


\date{\today}

\begin{abstract}
The excitation functions (EFs) for different fragments produced in the 
$^{27}$Al + $^{27}$Al dissipative collisions have been measured in steps  
of 250 keV in 
the incident energy range 122 -132 MeV. Deep inelastic processes have 
been selected by integrating events over a total kinetic energy loss 
(TKEL) window of 12 MeV between  20 and  32 MeV.    
Large fluctuations have been observed        
in all the studied EFs. Non-statistical origin of these fluctuations is 
confirmed by large channel cross correlation coefficients. Energy 
autocorrelation function (EAF) of the cross section EF presents damped 
oscillation structure as predicted for the case when   
a di-nuclear system  with a 
lifetime ($\tau$ = (5.7 $\pm$ 2.2)$\cdot$10$^{-21}$sec)  
similar with  its revolution period (T = 4.9$\cdot$10$^{-21}$sec) 
is formed.    The periodicity 
of the  EAF oscillations is used to obtain information 
on the deformation of the  
$^{27}$Al + $^{27}$Al di-nucleus.  
\end{abstract}

\pacs{25.,25.70.Lm,24.60.-k,24.60.Ky}

\keywords{}

\maketitle


\section{\label{sec:intro}Introduction}
Since 1985, when the presence of non-statistical fluctuations has been  
evidenced in the dependence of the cross section of deep inelastic 
processes on the incident energy \cite{rosa}, the excitation function 
statistical  analysis became a  method to obtain information on the di-nuclear 
system (DNS)  lifetime. Naturally it was addressed the question if the 
lifetime extracted from the EF analysis is identifiable with the value  
extracted 
 previously from the angular distributions of the final fragments. 
The authors of ref. \cite{bonetti} pointed out that the DNS 
lifetime extracted by these two methods are close to each other. They 
reached this conclusion based on a  reaction model in which  the  partial 
coherence mechanism  already addressed  
in \cite{abul,hartm} has been considered. Indeed the  model leads to an   
angular distribution pattern having a  degree of focusing given by the 
number of interfering partial waves  and  to 
fluctuating excitation functions. The analysis within this model has been done 
for the $^{28}$Si + $^{64}$Ni and  $^{12}$C + $^{24}$Mg systems  for which 
experimental information  was available by that time. Afterwards  
the excitations functions in dissipative heavy ion collisions (DHIC) have 
been measured and analyzed for other 
 light and  medium mass systems: $^{19}$F + $^{89}$Y, 
$^{28}$Si + $^{48}$Ti, $^{19}$F + $^{63}$Cu, 
$^{28}$Si + $^{28}$Si, $^{19}$F + $^{51}$V \cite{all}. 

A new model, Partial Overlapping Molecular Level Model  (POMLM), has 
been also developed in order to explain the presence of the 
fluctuations in the EF for DHIC  
introducing, besides the partial coherence, the hypothesis that the DNS  
is  excited in the low density region of molecular levels
 \cite{papa}.The large amplitude of the fluctuations persists due to the   
  lower density of the states in the vicinity of  the yrast line. 
  An unified explanation for fluctuation phenomenom in the excitation 
functions 
of  elastic, inelastic and deep inelastic collision has been obtained. 

The aim of the excitation functions for  $^{19}$F +  $^{27}$Al and 
$^{27}$Al + $^{27}$Al collisions   measured by 
us was to see to what extent the  fluctuation 
phenomenom is present in the case of these light systems where none of 
the participants to the reaction is an $\alpha$-conjugate nucleus.  

For the system $^{19}$F +  $^{27}$Al  
the excitation functions have been obtained for the final fragments 
with Z = 6 - 12  for a  
total kinetic energy loss (TKEL) window of 5 MeV width centered 
at 20 and 30 MeV in order  to study the dependence of this  phenomenom  
on TKEL. Such a study   has not been addressed before.
Non-statistical fluctuations have been evidenced in the cross section EFs of 
 final fragments and for  the corresponding fluctuation correlation 
widths, $\Gamma$, no dependence  on Z, TKEL and fragment emission angle, 
$\vartheta_{cm}$, has been observed within error limit \cite{ber}. 
An average value of $\Gamma$ of (170$\pm$65) keV has been obtained. 
The corresponding DNS lifetime  
(3.9 $\pm$ 1.1)$\cdot$10$^{-21}$ sec is in  agreement with the values 
extracted from angular distribution studies \cite{pop}. 

The present paper presents results obtained for the $^{27}$Al + $^{27}$Al 
collision.  
After a short review of the  experimental procedures and the criteria 
used to select 
deep inelastic events in Section \ref{expr}, results obtained 
from cross section EFs analysis   
for the $^{27}$Al + $^{27}$Al system are reported in subsection 
\ref{sa}  of the paper. The subsection \ref{evcor} is dedicated to the   
presentation of a study of the influence of the evaporation corrections on 
the fluctuating structure of EF. 
A pattern of the energy autocorrelation function  with  secondary damped 
oscillations besides the Lorentzian structure from $\varepsilon$ = 0, where 
$\varepsilon$ is the energy increment in the center of mass system (CMS), 
was obtained.  The observed oscillation periodicity 
 could be explained if hyperdeformed rotational states 
are supposed to be excited in DNS during  the collision \cite{kun3,kun4}. 
This aspect  is discussed in subsection \ref{deform}. Conclusions are 
presented in Section \ref{conclu}.

\section{\label{expr}Experimental procedures}

The experiment was performed at the SMP Tandem accelerator from LNS -Catania 
using $^{27}$Al ions with incident energies E$_{lab}$ = (122 - 132) MeV. 
 The energy loss  in   self 
supported targets of $^{27}$Al ($\approx$ 75 keV)  is less than  
the energy increment  of 250 keV (125 keV in CMS) 
between two successive experimental points.    
The beam 
current was measured with a tantalum plated Faraday cup provided with an 
electron suppressing  guard ring.

\begin{figure}[thb]
\scalebox{0.5}{\includegraphics{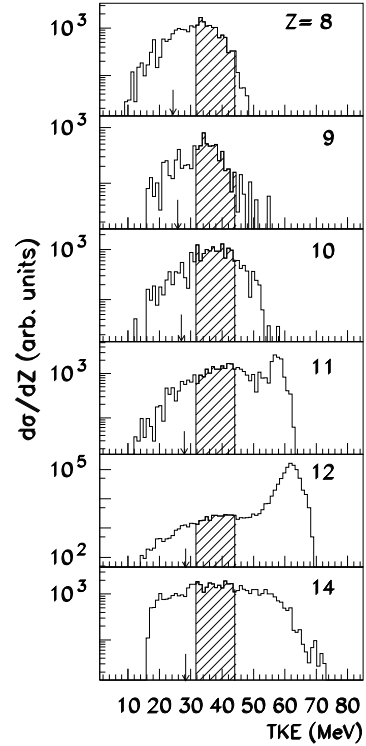}}
\caption{\label{fig1}Total kinetic energy spectra for Z = 8 - 12, 14 
reaction products 
in the $^{27}$Al + $^{27}$Al collision at incident energy E$_{lab}$ 
= 128 MeV.}
\end{figure}

The reaction products were detected and identified using the experimental 
device DRACULA \cite{petr,sim} 
 from which only the large area position sensitive ionization chambers (ICs) 
and the corresponding parallel plate avalanche counters in front of them 
have been operated. The ICs were filled with 
Ar(90\%) + CH$_{4}$(10\%) at 106.8 torr. The polar and azimuthal 
angles spanned by the IC  were $\Delta \vartheta$ = 24$^{\circ}$, 
$\Delta \varphi$ = 4$^{\circ}$, respectively. 
The energy resolution at the elastic peak was 2.5\%, the angular resolution 
 0.5$^{\circ}$ and the charge resolution better than 0.3 charge units.
The ICs being  centered at 24$^{\circ}$ in the laboratory system, an  angular 
range   12$^{\circ}\leq\vartheta_{lab}\leq36^{\circ}$ was covered 
continuously in a single measurement for each incident energy. The grazing 
angle at E$_{lab}$ =132 MeV is of $\approx$ 15 $^{\circ}$, so the weight 
of deep inelastic processes was increased within the measured angular 
interval. Nevertheless, a quasielastic 
component is present in the TKE spectra 
of fragments with Z = 11, 12 as could be seen in Fig.~\ref{fig1} for 
E$_{lab}$ = 128 MeV. 

The arrows 
on the figure represent the energies corresponding  to complete energy 
damping  \cite{viola}. The excitation functions for deep inelastic 
processes are obtained taking events with TKE larger than these values and 
well bellow the quasielastic component. The TKE  window used  
for E$_{lab}$ = 128 MeV is represented by the hatched area on  
Fig.~\ref{fig1} and corresponds to the  total kinetic energy loss window  
from 20 to 32 MeV, kept the same for all incident energies. 

\section{\label{crsecexf}Cross section excitation functions}

\subsection{\label{sa}Statistical analysis}

\begin{figure}[thb]
\scalebox{0.45}{\includegraphics{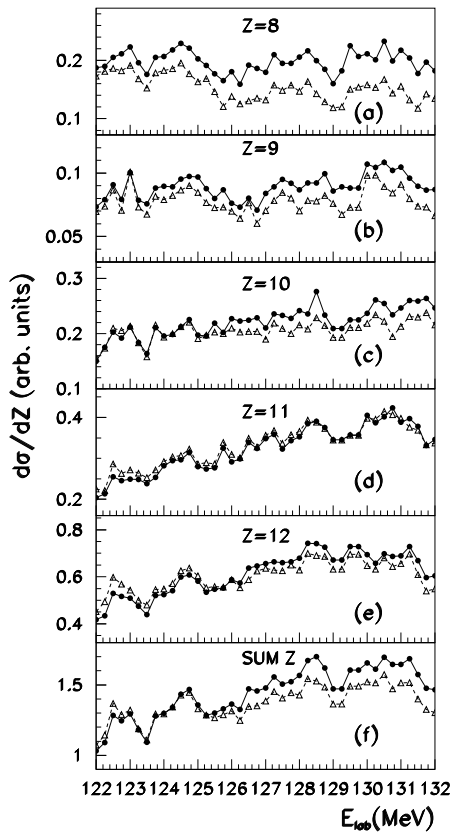}}
\caption{\label{fig2}Excitation functions for Z = 8 - 12 fragments 
with TKEL = 
(20 - 32) MeV produced $^{27}$Al + $^{27}$Al reaction (panels (a) - (e)), and 
summed EF of Z = 8, 11 and 12 products (panel (f)); the full circles  
represent the EFs 
after evaporation corrections (for primary products) and the empty triangles
 before their application
(for secondary products).}
\end{figure}

The obtained excitation functions present fluctuations with quite large 
amplitude as could be observed in Figs.~\ref{fig2},  where  the EFs 
corresponding to Z = 8 - 12, 14  are given. The full circles  represent the 
cross section of the primary products and the empty triangles of the secondary 
ones.

The influence of the evaporation corrections on the EFs, 
 obtained from 
comparing the results for these two sets of EF,
will be discussed  in  subsection \ref{evcor}. In  the following 
we refer only to the analysis of the EF for   primary products. 
  Fluctuations with amplitude larger than the statistical errors, 
which are of symbol size, could be observed. 
 The statistical analysis of the EFs has  
been done following the recipe from Ref. \cite{richter}.
Fluctuations present in the EFs corresponding to different final channels 
are strongly correlated, evidenced  by large channel cross correlation 
coefficients, $C_{Z_{i}Z_{j}}$, as can be seen  in Table ~\ref{tab1}. 
This  shows that 
the observed fluctuations are not of compound nucleus origin.  
Taking into account this correlation, in order to improve the statistics, 
 it was obtained an EF by summing the EFs for reaction products having  
higher yields (Z = 8,  11 and 12), which is given in the bottom panel
 of  Fig. ~\ref{fig2}. 
 
\begin{table}[b]
\caption{\label{tab1}Channel cross correlation coefficients}
\begin{ruledtabular}
\begin{tabular}{cccccc}
Z & 8  & 9 & 10 & 11 &12\\ 
8 & 1. & 0.865 & 0.702 & 0.356 & 0.467\\
9 &    &    1. & 0.482 & 0.671 & 0.554\\
10&    &       &1.     & 0.831 & 0.977\\
11&    &       &       &    1. & 0.703\\ 
12&    &       &       &       &     1.\\
\end{tabular}
\end{ruledtabular} 
\end{table}

\begin{figure}[thb]
\scalebox{0.55}{\includegraphics{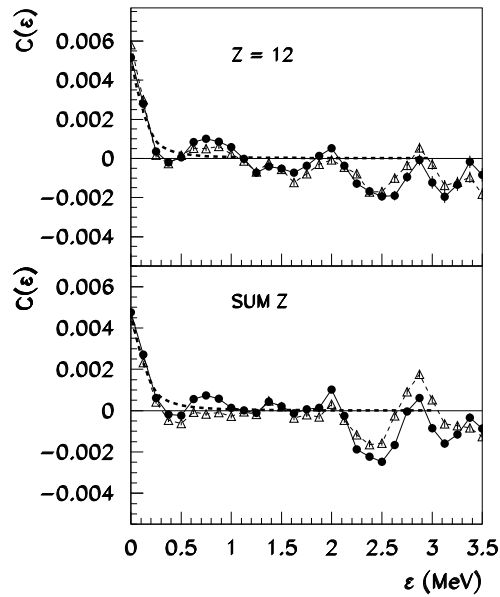}}
\caption{\label{fig3}Experimental energy autocorrelation functions 
 (points) and its fit with a 
Lorentzian function (dashed line); the full circles  and 
empty triangles  have the 
same meaning as in Fig. ~\ref{fig2}.}
\end{figure}

The experimental  energy autocorrelation functions  for 
the reaction product with the highest yield (Z=12) and for  
 summed EF are presented in Fig. ~\ref{fig3} by full circles.  
The average of the cross section over incoming energy has been  obtained 
using the  moving gauss  averaging procedure \cite{pappalardo}.  
The extracted fluctuation correlation widths $\Gamma_{p}$, by fitting  
the experimental EAFs with a Lorentzian function (thick dashed line in the 
same figure), are given in the first column of Table ~\ref{tab2}. The error of 
$\approx$ 25\% in the evaluation of $\Gamma_{p}$ 
is mainly due to finite range of data (FRD). For a   value of  
$\Gamma_{p}$ = (115$\pm$29) keV obtained for summed EF  corresponds  
a lifetime of the   $^{27}$Al +  $^{27}$Al DNS of 
(5.7 $\pm$ 2.2)$\cdot$10$^{-21}$sec,  specific for a DNS formed in the first 
stage of a heavy ion interaction.      

\begin{table}[b]
\caption{Energy correlation widths $\Gamma_{p}$ and $\Gamma_{s}$ for 
  primary and secondary products, respectively}
\label{tab2} 
\begin{ruledtabular}
\begin{tabular}{ccc}
Z & $\Gamma_{p}(keV)$ &  $\Gamma_{s}(keV)$ \\ 
8 &100$\pm$25& 113$\pm$29 \\
11&115$\pm$29&122$\pm$31  \\
12&104$\pm$26&115$\pm$29   \\
sum Z&115$\pm$29&122$\pm$31 \\
\end{tabular} 
\end{ruledtabular}
\end{table}

The  correlation width obtained by us is  
 very close  to  the value of $\Gamma$ =135 keV  
 obtained from the analysis of the  EF for  
dissipative collisions in the  $^{27}$Al + $^{27}$Al system on a lower 
incident energy interval E$_{lab}$ = 114 - 127 MeV in  Ref. \cite{wang}.

\begin{figure}[thb]
\scalebox{0.55}{\includegraphics{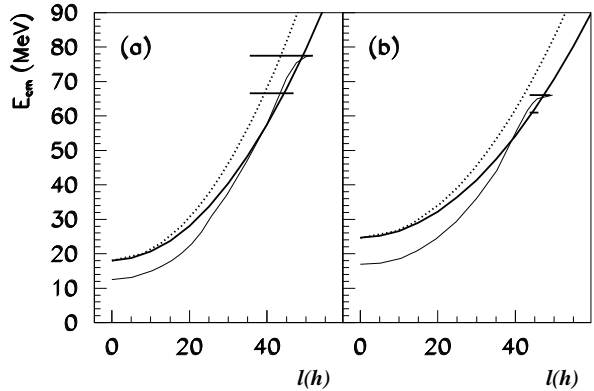}}
\caption{\label{fig4}Entry regions for (a) 
$^{19}$F + $^{27}$Al  system and (b) $^{27}$Al + $^{27}$Al system; horizontal 
lines correspond to the limits of the 
incident energy interval, the dotted line represents $l_{gr}$ as a function 
of 
incident energy, the thick continuous line - the OCM (yrast) line, the thin 
continuous line is the dependence of angular momentum as a function of TKE 
given by DONA code for the higher incident energy (Eq. (\ref{one})).}
\end{figure}

The  lower value of the correlation width obtained for the  
$^{27}$Al + $^{27}$Al system compared to  the case of  
 $^{19}$F + $^{27}$Al system    
  could be explained  by the lower excitation energy per nucleon 
available in heavier system. The position of the entry 
region (horizontal  lines correspond to upper and lower E$_{cm}$ energy 
for which measurements have been done) for   the two systems  
 is  given in  Fig.~\ref{fig4}. The entry regions for both 
systems lie in the vicinity of the yrast line (thick continuous line) 
given by the orbital cluster model (OCM) energies: 

\begin{equation}
\label{one}
E_{J} = V_{CB} + {\displaystyle\frac{J(J+1)}{2\cal{J}}},
\end{equation}    

\noindent
where $V_{CB}$ is the Coulomb barrier, $J$  the angular momentum 
and $\cal{J}$ the  total moment of inertia of di-nuclear system.

 The thin continuous line on the same figure  represents 
$l = f(TKE)$ as given by the calculation  based on the transport model 
(DONA code) \cite{wolshin} for the higher limit of the measured 
energy interval. One can see that the description of the $l $
dependence on $TKE$ within the transport model extensively used for heavy 
systems   works  quite well  for light 
systems too as it was already noticed   in our first study of a light system 
\cite{pop1}.

\subsection{\label{evcor}Influence of the evaporation corrections 
on cross section EF}

The extended theory of Ericson fluctuations and  the other models 
developed in order to explain 
the presence of non-statistical fluctuations in DHIC \cite{brink}  
  are elaborated for primary reaction products. 
Experimentally are measured the observables of secondary products 
and for comparison with models, evaporation corrections have to be done. 
The energy spectra were corrected for nucleon evaporation using 
the iterative procedure  described in Ref. \cite{breuer}. 
A parameterization of neutron and proton separation energies 
as a function of N/Z ratio elaborated for the mass region 
up to 50 has been used \cite{pop}. 
In order to estimate  the influence of these corrections on the EFs of 
the reaction products resulted from the $^{27}$Al + $^{27}$Al 
interaction,  we decided  
 to construct also a set of EFs before the  
evaporation corrections. These are represented with empty triangles in Fig.  
~\ref{fig2}. At the first sight  no sizable differences between the two sets 
of EFs could be seen. For a more quantitative 
evaluation, we calculated EAFs for EFs obtained without evaporation 
corrections. These are represented with empty triangles in Fig. ~\ref{fig3}. 
The extracted $\Gamma_{s}$ values are given in the second  
column of the Table ~\ref{tab2}. One can conclude that within error limits 
 a modification of  oscillation structure in the EFs the by evaporation 
corrections procedure is not observed, for the present data. 
It is not clear that the obtained result could be extended for higher  
excitation energy.     
   
\subsection{\label{deform}Deformation of di-nuclear system}

Experimental  EAF of DHIC shows, besides the Lorentzian structure at 
$\varepsilon$ = 0,   secondary oscillations  \cite{all,ber,wang}. 
 It was shown that  these oscillations appear when the lifetime of DNS 
is equal or greater than its rotation period  as a result of interference 
between different revolutions \cite{kun1,kun2}. The period of 
the secondary  structures is given by $\varepsilon = \hbar \omega  $, 
where $\omega$ is the 
angular velocity of the di-nucleus.

The experimental value of $\varepsilon$   in connection with the most 
probable kinetic energy $\langle$TKE$\rangle$ could be used to obtain 
an evaluation of the separation distance at the break-up moment 
 of the composite 
(di-nuclear) system and, in this way,  information on the deformation 
of the system during the interaction. 
  
The energy autocorrelation functions    from Fig.~\ref{fig3}  present   
secondary structures   with a period of $\approx$ 0.85 MeV. 
Such a period of oscillation implies for the distance of separation  
a value  of $r$  $\approx$ 10.9 fm, taking for the  
angular momentum the sticking configuration value given by  
$l_{st} = l(1-{\cal J}_{int}/{\cal J}_{tot})$, where $l = (l_{gr}+l_{cr})/2)$ 
and ${\cal J}_{int}, {\cal J}_{tot}$ are the intrinsic and total momenta of 
inertia, respectively. For $l_{gr}$ we used the value  of 47.8$\hbar$ 
calculated with the DONA code for $E_{lab}$ = 128 MeV, and 
$l_{cr}$ = 43.7$\hbar$  was calculated using  the formula given  
 in Ref. \cite{lcr}.   The value of the separation distance 
is compatibile with that  obtained by us   
 for  the $^{19}$F + $^{27}$Al system \cite{ber}. For this value of the  
separation distance   one obtains, for 
fragmentation $^{23}$Na + $^{31}$P,  
a value of $\langle$TKE$\rangle$ equal to  41.6 MeV. It was supposed that 
 $\langle$TKE$\rangle$ equals the final channel barrier given by the sum  of 
the Coulomb and centrifugal energies \cite{gelbke,nato}. 
The centrifugal energy is calculated    
for $l$ = $l_{st}$ $\approx$ 32.7 . This value is   in a  good agreement 
with the experimental most probable TKE of  $\approx$ 42 MeV   
 corresponding to  Z = 11 (see corresponding panel of  Fig. ~\ref{fig1}). 
One has to remark also that the used value of $l_{st}$ is within the 
$l$-window ($l$ = 27 - 33) corresponding to the integration   TKEL window 
(TKE = 32 - 44 MeV) given by the transport 
model calculation (thin continuous line in Fig.~\ref{fig4}(b)).      
The  separation distance values otained for both studied systems mean   
a quite large deformation for the reaction  partners of the  DNS. 
For relative momentum  of inertia at break-up,  the following relation can be 
written: 

\begin{equation}
\label{two}    
{\cal J}_{rel}(r) = 1.97\cdot {\cal J}_{rel}(R_{int}),
\end{equation}
 
\noindent
where $R_{int}$ = $r_{\circ}(A_{1}^{1/3}+A_{2}^{1/3})$ is the usual 
interaction radius. 

It was also necessary to  consider larger momenta 
of inertia in order to explain the period of 
secondary oscillations observed in the EAF of other systems 
($^{12}$C + $^{24}$Mg, 
$^{24}$Mg + $^{24}$Mg and $^{28}$Si + $^{28}$Si) in this mass 
region \cite{kun3,kun4}, while the period of secondary structures  of EAF 
for heavier systems ($^{19}$F +  $^{89}$Y, $^{58}$Ni + $^{58}$Ni, 
$^{58}$Ni +$^{62}$Ni, $^{58}$Ni + $^{46}$Ti) 
from Refs.  \cite{kun1,vanucci,kun5} could 
be explained using the moment of inertia calculated at $R_{int}$.      
 This shows the possibility of  the   existence of super- and 
hyper- deformation  in this 
mass region, as it was  predicted  thirty years ago  \cite{plasil}.  
The authors of Ref. \cite{plasil} concluded their paper:   
\textquotedblleft The outstanding problem is to devise methods that would 
identify 
the presence of such superdeformed nuclei which in some existing experiments 
have been produced without having been detected\textquotedblleft.    
Meanwhile experimental methods to evidence 
the existence of such nuclei have been developed. One should notice 
results of recent  experiments providing information on nuclear deformation  
in   the mass region close to the mass of the di-nuclear systems studied 
in the present paper, which appears to be a very interesting one from this 
point of view. 

For example  light charge particle spectra emitted by    
   $^{56}$Ni and  $^{44}$Ti compound nuclei could be described 
by using in formula (1) 
of the yrast line an effective moment of inertia ${\cal J}_{eff}$ = 
${\cal J}_{sphere}$(1+$\delta_{1}J^{2}+ \delta_{2}J^{4}$), where $\delta_{1}$, 
 $\delta_{2}$ are the deformability parameters and ${\cal J}_{sphere}$ is the  
rigid body moment of inertia \cite{alpha1,alpha2}. 
For deformability parameter values deduced from the fit 
of experimental spectra, one obtaines for the effective moment of inertia 
${\cal J}_{eff}$ = 1.4${\cal J}_{sphere}$ and ${\cal J}_{eff}$ = 
1.5${\cal J}_{sphere}$ 
  for the $^{56}$Ni and $^{44}$Ti CN, respectively.    
Eq. (\ref{two}) obtained for the di-nucleus $^{54}$Fe is  
similar to these relations. The relative momentum of 
inertia at break-up obtained by us can be also  seen  as an effective 
momentum introduced in order to reproduce the experimental periodicity of 
EAF.

The  $\gamma$-ray spectrum from the decay of GDR built in hot 
$^{46}$Ti was recently measured and its shape suggests the presence of an  
elongated 3-axial equilibrium shape for this nucleus at J=30$\hbar$ 
\cite{maj1,maj2}. The excitation functions for the system 
$^{19}$F +  $^{27}$Al  
(di-nucleus $^{46}$Ti) have been constructed for TKEL windows 
centered at 20 and 30 MeV. One can see from Fig. ~\ref{fig4}(a) that for these 
TKEL windows (TKE = 50 - 55 MeV and 40 - 45 MeV, respectively) 
correspond angular momentum values near the region where very elongated 
shapes could appear.

 Thus, one can consider the analysis of the cross section 
EFs as a robust experimental method to evidence and extract  information 
on such elongated objects. 
The results obtained in the present paper and in the previous 
ones \cite{ber,kun3,kun4}, using this method,
 evidence a large deformation of composite systems 
formed in the first stage of elastic, inelastic scatering and deep 
inelastic   processes  in the mass region 36 - 56.

\section{\label{conclu}Conclusions}

The EFs analysis for deep inelastic processes 
in the $^{27}$Al + $^{27}$Al collision  shows that rotational 
states with large deformation are excited  in the first stage 
of the reaction. The correlation width of the fluctuations extracted from 
cross section EF is equal to (115$\pm$29) keV   to which coresponds a 
DNS lifetime of  (5.7 $\pm$ 2.2)$\cdot$10$^{-21}$sec. 
The large channel cross correlation coefficients show the non-statistical 
origin of the fluctuations.  
They support  
a reaction mechanism where  special states 
of rotational (molecular) nature play the role of doorway configurations 
towards a regime characterized by stochastic exchange of nucleons 
between interacting nuclei as the main mechanism  behind the dissipative 
phneomena in light heavy ion collisions.

\begin{acknowledgments}
We are grateful to the operating crew of LNS Tandem accelerator for 
the quality of delivered beam.  
For the high quality targets we owe special acknowledgement  
to Mr. C. Marchetta. We acknowledge Prof. E. Migneco and   Prof. D.  
Vinciguerra for their permanent support.   
\end{acknowledgments}


\end{document}